\documentclass[preprint,showpacs,preprintnumbers,amsmath,amssymb,superscriptaddress]{revtex4}

\include{ams}
\usepackage{amsmath,amssymb,amsthm}
\usepackage{dcolumn}
\usepackage{bm}
\usepackage{graphicx}

\begin{document}

\title{Spheroidal geometry approach to fullerene molecules}

%\author{1st author name}
%\altaffiliation[Also at ]{Dept. of physics, FMFI Comenius University, Bratislava}
%\email{1st@uniba.sk}
%
%\author{2nd author name}
%\email{2nd@savba.sk}
%\affiliation{Institute of Physics, Slovak Academy of Sciences, Bratislava}
%
%\author{3rd author name}
%\email{3rd@jinr.ru}
%\affiliation{Joint Institute for Nuclear Research, Dubna, Russia}

\author{R. Pincak}
\email{pincak@saske.sk}
\altaffiliation[on leave from ]{Institute of Experimental Physics, Slovak Academy of Sciences, Watsonova 47,043 53 Kosice, Slovak Republic}
\affiliation{Joint Institute for Nuclear Research, BLTP, 141980 Dubna, Moscow region, Russia}


\date{\today}

\begin{abstract}
Graphite is an example of a layered material that can be bent to
form fullerenes which promise important applications in electronic
nanodevices. The spheroidal geometry of a slightly elliptically
deformed sphere was used as a possible approach to fullerenes. We
assumed that for a small deformation the eccentricity of the
spheroid $e\ll 1$. We are interested in the elliptically deformed
fullerenes C$_{70}$ as well as in C$_{60}$ and its spherical
generalizations like big C$_{240}$ and C$_{540}$ molecules. The
low-lying electronic levels are described by the Dirac equation in
(2+1) dimensions. We show how a small deformation of spherical
geometry evokes a shift of the electronic spectra compared to the
sphere. The flux of a monopole field was included inside the surface
to describe the fullerenes. Both the electronic spectrum of
spherical and the shift of spheroidal fullerenes were derived.

\end{abstract}

\maketitle

\section{Introduction}

Fullerene molecules~\cite{1} are carbon cages which appear in the
vaporization of graphite. One of their most beautiful features from
a formal point of view is their geometric character and the exciting
possibility of producing them in all sorts of geometric shapes
having as building blocks sections of the honeycomb graphite
lattice. The most abundant of them is the most spherical $C_{60}$
molecule. The shape of the $C_{60}$ molecule is that of a soccer
ball, consisting of 12 pentagons and 20 hexagons. However, some
fullerenes as $C_{70}$ are slightly elliptically deformed with the
shape being more similar to an American football. Fullerenes belong
to a sort of carbon nanoparticles.

Carbon nanoparticles, which are expected to have important
implications for the development of electronic devices, flat panel
displays, nano-switches, etc., have recently received great
attention of both experimentalists and theorists (see~\cite{2}).
High flexibility of carbon allows one to produce variously shaped
carbon nanoparticles: nanotubes, fullerenes, cones, toroids,
graphitic onions and nanohorns. Particular attention was given to
peculiar electronic states due to topological defects which were
observed in different kinds of carbon nanoparticles by scanning
tunneling microscopy (STM). For example, STM images with five-fold
symmetry (due to pentagons in the hexagonal graphitic network) were
obtained in the C$_{60}$ fullerene molecule~\cite{3}. The peculiar
electronic properties at the ends of carbon nanotubes (which include
several pentagons) were probed experimentally in~\cite{4,5}.

By its nature, the pentagon in a graphite sheet is a topological
defect. Actually, as was mentioned in Ref.~\cite{6}, fivefold
coordinated particles are orientational disclination defects in the
otherwise sixfold coordinated triangular lattice. The local density
of states was found in the vicinity of a pentagonal defect for
spherical fullerenes~\cite{7,8}. Moreover, disclinations are {\it
generic} defects in closed carbon structures, fullerenes and
nanotubes, because, in accordance with Euler's theorem, these
microcrystals can only be formed by having a total disclination of
$4\pi$. According to the geometry of the hexagonal network, this
implies the presence of twelve pentagons ($60^{\circ}$
disclinations) on the closed hexatic surface.

Investigation of the electronic structure requires formulating a
theoretical model describing electrons on arbitrary curved surfaces
with disclinations taken into account. An important ingredient of
this model can be provided by the self-consistent effective-mass
theory describing the electron dynamics in the vicinity of an
impurity in graphite intercalation compounds~\cite{9}. The most
important fact found in~\cite{9} is that the electronic spectrum of
a single graphite plane linearized around the corners of the
hexagonal Brillouin zone coincides with that of the Dirac equation
in (2+1) dimensions. This finding stimulated a formulation of some
field-theory models for Dirac fermions on hexatic surfaces to
describe electronic structure of variously shaped carbon materials:
fullerenes~\cite{10,11} and nanotubes~\cite{12}.

The Dirac equation for massless fermions in three-dimensional
space-time in the presence of the magnetic field was found to yield
$N-1$ zero modes in the N-vortex background field~\cite{13}. As was
shown in Ref.~\cite{14}, the problem of the local electronic
structure of fullerene is closely related to Jackiw's
analysis~\cite{13}. Notice that the field-theory models for Dirac
fermions on a plane and on a sphere~\cite{15} were invoked to
describe variously shaped carbon materials. Recently, the importance
of the fermion zero modes was discussed in the context of
high-temperature chiral superconductors and fullerene molecules.

The most spherical fullerene is the C$_{60}$ molecule nicknamed a
'bucky ball'. Others are either slightly (as C$_{70}$ whose shape is
more like an elliptic deformation) or remarkably deformed. We are
interested here in the C$_{60}$ molecule as well as in its spherical
generalizations like big C$_{240}$ and C$_{540}$ molecules with the
symmetry group of the icosahedron, and also in the elliptically
deformed fullerene C$_{70}$ and its relatives. Big fullerenes are
used to store radioactive material and inhibit enzymes related to
different viruses~\cite{16,17}.

\section{The model}

Almost all fullerenes are only slightly elliptically deformed
spherical molecules, e.g., C$_{70}$ and its relatives. We start with
introducing spheroidal coordinates and writing down the Dirac
operator for free massless fermions on the Riemann spheroid $S^{2}$.
Pi-molecular  orbitals in fullerenes as a free electron model
(electron gas) bound on the surface of a sphere were used
in~\cite{18}. We generalize that work to obtain an electronic
spectrum for spherical and spheroidal geometries with and without
the monopole field. The peculiarities of the electronic spectra for
these two slightly different types of geometries are shown.

To incorporate fermions on the curved background, we need a set of
orthonormal frames $\{e_{\alpha}\}$, which yield the same metric,
$g_{\mu\nu}$, related to each other by the local $SO(2)$ rotation,
$$e_{\alpha}\to e'_{\alpha}={\Lambda}_{\alpha}^{\beta}e_{\beta},\quad
{\Lambda}_{\alpha}^{\beta}\in SO(2).$$ It then follows that
$g_{\mu\nu} = e^{\alpha}_{\mu}e^{\beta}_{\nu} \delta_{\alpha \beta}$
where $e_{\alpha}^{\mu}$ is the zweibein, with the orthonormal frame
indices being $\alpha,\beta=\{1,2\}$, and the coordinate indices
$\mu,\nu=\{1,2\}$. As usual, to ensure that physical observables are
independent of a particular choice of the zweibein fields, a local
$so(2)$ valued gauge field $\omega_{\mu}$ is to be introduced. The
gauge field of the local Lorentz group is known as a spin
connection. For a theory to be self-consistent, the zweibein fields
must be chosen to be covariantly constant \cite{19}
$${\cal D}_{\mu}e^{\alpha}_{\nu}:=\partial_{\mu}e^{\alpha}_{\nu}
-\Gamma^{\lambda}_{\mu\nu}e^{\alpha}_{\lambda}+(\omega_{\mu})^{\alpha}_{\beta}
e^{\beta}_{\nu}=0,$$ which determines the spin connection
coefficients explicitly
\begin{equation}
(\omega_{\mu})^{\alpha\beta}= e_{\nu}^{\alpha}D_{\mu}e^{\beta\nu}.
\label{eq:1}\end{equation}

Finally, the Dirac equation on a surface $\Sigma$ in the presence of
the magnetic monopole field $A_{\mu}$ is written as~\cite{20}
\begin{equation}
i\gamma^{\alpha}e_{\alpha}^{\ \mu}[\nabla_{\mu} -
iA_{\mu}]\psi=E\psi, \label{eq:2}\end{equation} where
$\nabla_{\mu}=\partial_{\mu}+\Omega_{\mu}$ with
\begin{equation}
\Omega_{\mu}=\frac{1}{8}\omega^{\alpha\ \beta}_{\ \mu}
[\gamma_{\alpha},\gamma_{\beta}] \label{eq:3},\end{equation} being
the spin connection term in the spinor representation.

The elliptically deformed sphere or a spheroid
\begin{equation}
\frac{x^{2}}{a^{2}}+\frac{y^{2}}{a^{2}} +\frac{z^{2}}{c^{2}}=1,
\label{eq:4}
\end{equation}
may be parameterized by two spherical angles $q^{1}=\theta$,
$q^{2}=\phi$ that are related to the Cartesian coordinates $x$, $y$,
$z$ as follows
\begin{equation}
x=a~\sin\theta \cos\phi; \quad y=a~\sin\theta \sin\phi; \quad
z=c~\cos\theta. \label{eq:5}\end{equation} We have assumed that the
eccentricity of the spheroid is $e\ll 1$ which in the case $c<a$
gives expressions $e=\sqrt{1-(\frac{c}{a})^{2}}\ll 1$. The metric
tensor and the natural diagonal zweibein different from zero on the
spheroid are
\begin{equation}
g_{\phi\phi}=a^{2}\sin^{2}\theta; \quad
g_{\theta\theta}=a^{2}\cos^{2}\theta+c^{2}\sin^{2}\theta, \quad
\label{eq:6}\end{equation} where~$a,c\geq0,~0\leq\theta\leq\pi,~
0\leq\phi<2\pi$ and
\begin{equation}
e{_{\phi}^{2}}=\frac{1}{a~\sin\theta};\quad
e{_{\theta}^{1}}=\frac{1}{\sqrt{a^{2}\cos^{2}\theta+c^{2}\sin^{2}\theta}}\label{eq:7}
~,\end{equation} which, in view of Eq.~(\ref{eq:1}) gives the spin
connection coefficients
\begin{equation}
\omega{_{\phi}^{12}}=-\omega{_{\phi}^{21}}=\frac{a}{\sqrt{a^{2}+c^{2}\tan^{2}\theta}}=:2\omega.
\label{eq:8}\end{equation} In $2D$ the Dirac matrices can be chosen
to be the Pauli matrices, $\gamma^1=-\sigma^2, \gamma^2=\sigma^1$;
Eq.~(\ref{eq:3}) then reduces to
\begin{equation}
\Omega_{\phi}=i\omega\sigma^3.\label{eq:9} \end{equation} We have
assumed that $A_{\theta}=0$ and only the monopole field $A_{\phi}$
is different from zero.

The eigenfunctions of the Dirac operator are two-component spinors
that satisfy the eigenvalue equation
\begin{equation}
-i\widehat{\nabla}\left(\alpha_{\lambda}(\theta,\phi)\atop{\beta_{\lambda}(\theta,\phi)}\right)
=\lambda\left(\alpha_{\lambda}(\theta,\phi)\atop{\beta_{\lambda}(\theta,\phi)}\right)
.\label{eq:10}
\end{equation}
%where $\lambda=ER$ and $R$ is the radius of the spheroid.
This system of first order partial differential equations in
$\alpha$ and $\beta$ allows separation of variables therefore we can
isolate the $\phi$-dependence by expanding the spinors into the
Fourier series
\begin{equation}
\Psi(\theta,\phi)=\left(\alpha_{\lambda}(\theta,\phi)\atop{\beta_{\lambda}(\theta,\phi)}\right)
=\sum_{m}\frac{\exp~im\phi }{\sqrt{2\pi}}\left(\alpha_{\lambda m
}(\theta)\atop{\beta_{\lambda m }(\theta)}\right);\quad
m=\pm\frac{1}{2},\pm\frac{3}{2}..., \label{eq:11}
\end{equation}
where $m$ are half-integers since we work with the spin
$\frac{1}{2}$~field.
%Firstly we will assume that field $A_{\phi}=0$.
Then the general form of the Dirac equation, Eq.~(\ref{eq:2}), on
the spheroid becomes
%\begin{subequations}\label{eq:1.5}
$$
\bigg[\partial_{\theta}+\sqrt{\cot^{2}\theta+\bigg(\frac{c}{a}\bigg)^{2}}m+\bigg(1-2A_{\phi}\sqrt{1+\bigg(\frac{c}{a}\bigg)^{2}\tan^{2}\theta}\bigg)~
\frac{\cot\theta}{2}\bigg]\beta_{\lambda m }(\theta)= \nonumber
$$
$$
-\sqrt{a^{2}\cos^{2}\theta+c^{2}\sin^{2}\theta}~E\alpha_{\lambda
m}(\theta),\nonumber
$$
%\end{subequations}
$$
\bigg[\partial_{\theta}-\sqrt{\cot^{2}\theta+\bigg(\frac{c}{a}\bigg)^{2}}m+\bigg(1+2A_{\phi}\sqrt{1+\bigg(\frac{c}{a}\bigg)^{2}\tan^{2}\theta}\bigg)~
\frac{\cot\theta}{2}\bigg]\alpha_{\lambda m }(\theta)= \nonumber
$$
\begin{equation}
\sqrt{a^{2}\cos^{2}\theta+c^{2}\sin^{2}\theta}~E\beta_{\lambda
m}(\theta) \label{eq:12}. \end{equation} The number $m$ may be
called the projection of angular momentum onto the polar axis. If
$a=c=R$, where $R$ is the radius of a sphere, Eq.~(\ref{eq:12})
becomes the Dirac equation for sphere geometry.

Now we want to find an electronic spectrum for the sphere and
spheroid analytically and numerically, respectively; therefore, we
firstly assume that pentagon defects represented in this model by
the monopole field $A_{\phi}=0$. Next, we want to separate the
equations for the spinor components $\alpha$ and $\beta$. This can
be done by taking the square $\Delta$ (Laplace operator) of the
Dirac operator $\hat{\nabla}$ for spheroidal geometry. Finally, we
find the equations
$$
\bigg[-\frac{1}{\sin\theta}~\partial_{\theta}~\sin\theta~\partial_{\theta}+\bigg(\cot^{2}\theta+\bigg(\frac{c}{a}\bigg)^{2}\bigg)m^{2}-\sigma^{3}\frac{\cot\theta}{\sqrt{\cot^{2}\theta+\bigg(\frac{c}{a}\bigg)^{2}}}\frac{m}{\sin^{2}\theta}+\frac{1}{4
\sin^{2}\theta}\bigg]\left(\alpha_{\lambda m
}(\theta)\atop{\beta_{\lambda m }(\theta)}\right)=
$$
\begin{equation}
\bigg[(a^{2}\cos^{2}\theta+c^{2}\sin^{2}\theta)~E^{2}-\frac{1}{4}\bigg]\left(\alpha_{\lambda
m }(\theta)\atop{\beta_{\lambda m }(\theta)}\right)
\label{eq:13}.\end{equation} Further simplifications come from the
change of variables $x=\cos\theta$, $x\in[-1,1]$, which converts
Eq.~(\ref{eq:13}) into the generalized hypergeometric equations
$$
\bigg[\frac{d}{dx}(1-x^{2})\frac{d}{dx}-\bigg(\frac{x^{2}}{1-x^{2}}+\bigg(\frac{c}{a}\bigg)^{2}\bigg)m^{2}+\sigma^{3}\frac{x}{\sqrt{x^{2}+\bigg(\frac{c}{a}\bigg)^{2}(1-x^{2})}}\frac{m}{1-x^{2}}-\frac{1}{4(1-x^{2})}
\bigg]\left(\alpha_{\lambda m }(x)\atop{\beta_{\lambda
m}(x)}\right)=
$$
\begin{equation}
-\bigg[(a^{2}x^{2}+c^{2}(1-x^{2}))~E^{2}-1/4\bigg]\left(\alpha_{\lambda
m }(x)\atop{\beta_{\lambda m }(x)}\right)
\label{eq:14}.\end{equation} The replacement $x\longrightarrow-x$
(or $m\longrightarrow-m$) is equivalent to changing $\alpha$ for
$\beta$. Thus, the upper and lower spinor components are conjugate
with respect to mirror reflection. Equation~(\ref{eq:14}) is
singular at the poles of the spheroid $x=\pm1$. We redefine the
unknowns
\begin{equation}
\left(\alpha_{\lambda m }(x)\atop{\beta_{\lambda m
}(x)}\right)=\left((1-x)^{1/2|m-1/2|}(1+x)^{1/2|m+1/2|}\xi_{\lambda
m} (x) \atop{(1-x)^{1/2|m+1/2|}(1+x)^{1/2|m-1/2|}\eta_{\lambda m}
(x)}\right) ,\label{eq:15}\end{equation} and use that in our model
of a slightly deformed sphere $c\sim a$; hence
$\frac{c}{a}\doteq1\pm\delta$, where $\delta\ll1$ is small
deformation of the sphere. Neglecting the second and higher order
powers of $\delta$ we solve Eq.~(\ref{eq:14}) to the first order in
$\delta$ and by using redefinition (\ref{eq:15}) we arrive at the
separate equations of hypergeometric type in $\xi_{\lambda m}$ and
$\eta_{\lambda m}$
\begin{equation}
\bigg\{(1-x^{2})\frac{d^{2}}{dx^{2}}-(1-2(m-1)x)\frac{d}{dx}-m(m-1)+\lambda^{2}-1/4+f(x)\bigg\}\xi_{\lambda
m }(x)=0 ,\label{eq:16}
\end{equation}
where
$$
f(x)=\delta(\mp2m^{2}\pm2 a^{2}E^{2}\mp m x\mp 2
a^{2}E^{2}x^{2});\quad \lambda=Ea .
$$
For the above calculations we have assumed that $m-1/2\leq0$. A
similar solution can be also found for the case $m-1/2\geq0$.
Following the calculations above we can get also the equation for
function $\eta_{\lambda m} (x)$. The function $f(x)$ is the
deviation of the solution for a spheroid from that for a sphere,
see~\cite{21}, and can be perceived as next energy part, i.e. the
energy shift for a spheroid compared to a sphere geometry. Thus the
expression $f(x)$ can also be called as a perturbation part of
Equations~(\ref{eq:16}).
%The expression for the local maximum and minimum
%values of the deviation with $\delta\neq0$ can be found analytically
%from the equation $f'(x)=0$ in the form $x=-m/4 a^{2} E^{2}$. The
%analytical expression for the maximum/minimum values of the
%deviation (shift) has the following form: $f_{max/min}(x)=\pm 2
%a^{2}E^{2}\delta\mp 2m^{2}\delta\pm m^{2}\delta/8 a^{2}E^{2}$ which
%depends on the choice of the signs in $f(x)$.

To find the spectrum for spherical geometry $a=c=R$, we have to put
the expression $f(x)=0$, so the case if $\delta=0$. The spectrum can
be found in the form
\begin{equation}
\lambda^{2}_{sphere}=(n+|m|+1/2)^{2},\label{eq:17}
\end{equation}
with non-negative integer $n\geq0$, with $n$ being the order of
Jacobi polynomials, see~\cite{21}. The eigenvalues $\lambda$ for the
sphere of the unit radius $S^{2}$ are nonzero integers
\begin{equation}
\lambda_{sphere}=\pm1,\pm2..., \label{eq:18}
\end{equation}
and indeed the Dirac operator has no zero-modes. The spectra of the
spherical geometry as the numerical calculations of
Eq.~(\ref{eq:12}) with $A_{\phi}=0$ and for $a=c$ are illustrated in
Fig.~1 and fit the analytical results in (18).

\begin{figure}[htb]
  \centering
\includegraphics[width=0.37\textwidth]{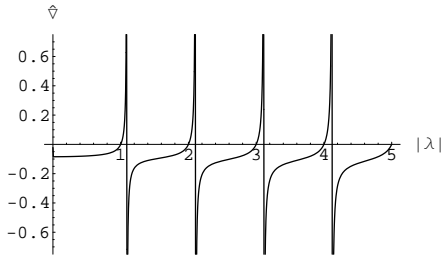}
\caption{ The electronic spectra of spherical geometry
$|\lambda|_{sphere}$, where $a=c$.}
\end{figure}

To find the electronic spectrum also in the case of spheroidal
geometry when $f(x)\neq0$, we solve Eq.~(\ref{eq:12}) for two cases,
when $a<c$ and $a>c$, i.e. for two different types of an
elliptically deformed sphere. The numerical results are shown in
Fig.~2.

\begin{figure}[htb]
  \centering
\includegraphics[width=0.9\textwidth]{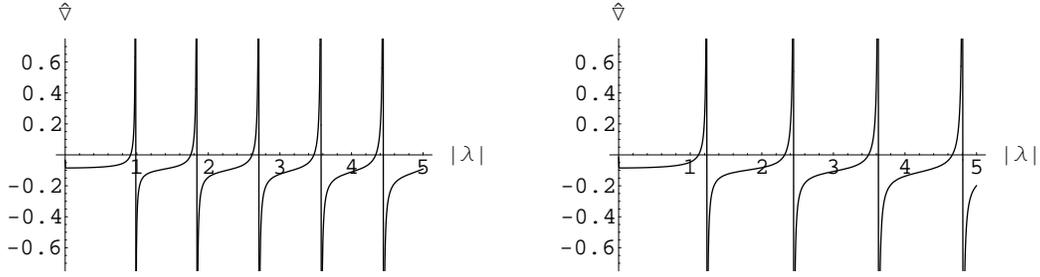}
\caption{ The electronic spectra of spheroidal geometry
$|\lambda|_{spheroid}$ where, $a<c$ and $a>c$ (going from left to
right).}
\end{figure}

%The perturbation parameter $\delta=0.3$ was used to obtain the
%electronic spectrum of the Dirac equation for spheroidal geometry.
As can be seen from Fig.~2, the spectra display the energetic shifts
compared to the spectra of the sphere in Fig.~1. The shift increases
or reduces the initial electronic spectrum of sphere depending on
the type of elliptic deformation. The shift is bigger with
increasing value of the modes of the electronic spectra, which can
be also seen from the structure of $f(x)$ in Eq.~(\ref{eq:16}). The
perturbation parameter $\delta=0.3$ was used to obtain the
electronic spectra of the Dirac equation for the spheroidal
geometry. The angular momentum $m=1/2$ was used in the calculations.
All the spectra are mirror symmetric with respect to the $y$-axis.

\section[]{Dirac equations for spheroid with monopole field}

We assume that the parameter of perturbation and the eccentricity of
the spheroid $e\ll1$, so we can use the magnetic monopole field
inside the surface to obtain $C_{70}$ fullerene or bigger fullerene
molecules like C$_{240}$ and C$_{540}$ also with a small elliptical
deformation. The area of surface for a small elliptically deformed
sphere, spheroid, e.g. for oblate spheroid ($a=b$) can be formulated
as, see~\cite{22}
$$
S\approx2\pi(a^{2}+c^{2}\frac{\textrm{arctanh}(e)}{e})\nonumber.
$$
In the case of small eccentricity the surface area of a spheroid
becomes the surface area of a sphere $S\approx2\pi a^{2}$. It means
that we can include the existence of a fictitious magnetic monopole
charge $g$ inside the surface of the spheroid with the structure as
in Ref.~\cite{15}. The values of $g$, e.g. for tetrahedron and
icosahedron structures required $1/2$ and $3/2$, respectively. With
the monopole field taken as $A_{\theta}=0$ and
$A_{\phi}=\frac{j}{2}\cos\theta$, where $j/2=g$ (for structures
above $j=1,3$), the resulting Dirac equation~(\ref{eq:12}) for the
spheroid with the monopole field reads
$$
\bigg[\partial_{\theta}+\sqrt{\cot^{2}\theta+\bigg(\frac{c}{a}\bigg)^{2}}m+\bigg(1-\frac{j}{a}\sqrt{a^{2}\cos^{2}\theta+c^{2}\sin^{2}\theta}\bigg)~
\frac{\cot\theta}{2}\bigg]\beta_{\lambda m }(\theta)= \nonumber
$$
$$
-\sqrt{a^{2}\cos^{2}\theta+c^{2}\sin^{2}\theta}~E\alpha_{\lambda
m}(\theta)\nonumber,
$$
%\end{subequations}
$$
\bigg[\partial_{\theta}-\sqrt{\cot^{2}\theta+\bigg(\frac{c}{a}\bigg)^{2}}m+\bigg(1+\frac{j}{a}\sqrt{a^{2}\cos^{2}\theta+c^{2}\sin^{2}\theta}\bigg)
~ \frac{\cot\theta}{2}\bigg]\alpha_{\lambda m }(\theta)= \nonumber
$$
\begin{equation}
\sqrt{a^{2}\cos^{2}\theta+c^{2}\sin^{2}\theta}~E\beta_{\lambda
m}(\theta) \label{eq:19}.\end{equation} Unfortunately, general
solutions to Eq.~(\ref{eq:19}) are not available yet for spherical
and spheroidal fullerenes, so we do not have initial conditions for
numerical calculations which are very sensitive to them. However, we
will present some analytical predictions of the electronic spectra
for spherical and spheroidal fullerenes from the square of Dirac
operator in Eq.~(\ref{eq:19}).

To find an analytical expression for the shift of the spheroidal
fullerenes, we put for simplicity in Eq.~(\ref{eq:19}) the value of
the angular momentum $m=0$. For elliptically deformed fullerenes in
the special case where $m=0$ with using the model of a slightly
deformed sphere where $\frac{c}{a}\doteq1\pm\delta$, the square of
Dirac operator (19) was found in the form
$$
\bigg[-\frac{1}{\sin\theta}~\partial_{\theta}~\sin\theta~\partial_{\theta}+\frac{(j\sqrt{(1\pm2\delta\sin^{2}\theta)}-\sigma^{3})^{2}}{4\sin^{2}\theta}+\sigma^{3}\frac{\pm
j\delta\cos^{2}\theta}{\sqrt{(1\pm2\delta\sin^{2}\theta)}}\bigg]\left(\alpha_{\lambda
m }(\theta)\atop{\beta_{\lambda m }(\theta)}\right)
$$
\begin{equation}
=\bigg[(\lambda^{2}+(j/2)^{2})(1\pm2\delta\sin^{2}\theta)-1/4\bigg]\left(\alpha_{\lambda
m }(\theta)\atop{\beta_{\lambda m }(\theta)}\right)
\label{eq:20},\end{equation} where $\delta\ll1$ is small deformation
of the sphere. We neglect the second and higher order powers of
$\delta$ in the calculations.

If the perturbation parameter $\delta=0$, we get the square of Dirac
operator for spherical fullerenes with the electronic spectrum as in
Ref.~\cite{11}. To get connection with Section II, the spectrum with
the monopole field for the spherical fullerenes can be rewrite from
the right hand side of Eq.~(\ref{eq:20}) to the form
\begin{equation}
\lambda^{2}_{field}=(|\lambda|_{sphere}-1/2)^{2}-g^{2},\label{eq:21}
\end{equation}
with $|\lambda|_{sphere}-1/2=n+|m|$, see~\cite{21}. The value of the
electronic spectrum $\lambda_{field}$ is shifted (decreases) by the
value of the charge $g$ of the monopole field.
%Here $l$ is the total angular momentum and it can get the values
%$l=1/2,3/2,5/2...$. The choice of the angular momentum in the
%calculations depends on the choice of the spinor representations. In
%our model the eigenfunctions are two-component spinors that belong
%to representations of the $SU(2)$-group with half-integer angular
%momenta $l$.
Moreover, the presence of the monopole restricts possible values of
the angular momentum~\cite{15}, so that $m\geq\mid\mid j\mid-1\mid/2
$ and, therefore, the values of $m$ change contrary to structures
without a monopole field inside, as in Section II. It means that for
fullerenes the angular momentum $m$ can get the values $m=0,1...$.
The spectra of fullerenes are appended by the monopole charge
compared to Eq.~(\ref{eq:13}) for spherical geometry. So the spectra
of fullerenes are greatly dependent on the value of the monopole
field.
%Moreover, there is a minimum value of the angular momentum
%dictated by the charge $g$; therefore, the number of zero modes in
%the spectrum also depends exclusively on the value of the monopole
%charge.
In the case of tetrahedron and icosahedron structures expression
(21) can predict the existence of zero modes where
$\lambda_{field}=0$.

If $\delta\neq0$, the shift of the spheroidal fullerenes from the
spherical ones in Eq.~(\ref{eq:20}) were found. Moreover, when we
use the substitution $x=\cos\theta$ as in Section II, we can rewrite
the shift on the right hand side of Eq.~(\ref{eq:20}) in the
following form
\begin{equation}
f(x)=2\delta(\lambda^{2}+g^{2})(\pm1 \mp x^{2});\quad \lambda=a E.
\end{equation}\label{eq:22}
The electronic spectrum will be shifted by the value
$\pm2\delta(\lambda^{2}+g^{2})$. The function $f(x)$ is the
deviation of spheroidal from spherical fullerenes, similarly to the
previous section, and can also be perceived as the energy shift to
spheroidal fullerenes compared to spherical ones. Moreover, the
perturbation function $f(x)$ is in the case when $m=0$ and $g=0$ the
same as the shift in Eq.~(\ref{eq:16}) for spheroidal geometry. So
we can expect a similar behavior (shifts) of the spectra of
spheroidal fullerenes (decreases by the value of the monopole charge
$g$) as in the case of spheroidal geometry without a monopole field,
see Fig.~2.

\section{Conclusion}

To find the electronic spectra of the $C_{70}$ fullerene and its
relatives, we have used the model of a slightly elliptically
deformed sphere, the spheroidal geometry ($\delta\neq0$), as
distinct from the $C_{60}$ fullerene where the spherical geometry
($\delta=0$) was used. The Dirac equation in (2+1) dimensions for
slightly elliptically deformed fullerenes with monopole field inside
the surface was evaluated. The discrete spectrum of energy for both
types of geometries was found. The electronic spectrum and the shift
of the spheroidal geometry in Eq.~(\ref{eq:16}) contrary to the
sphere was calculated both analytically and numerically. Figure~2
shows the shift of the spectra for spheroidal geometry. In the case
of spherical and spheroidal fullerenes the electronic spectrum and
shift were derived analytically. The spectra of spherical and
spheroidal fullerenes decreasing by the value of monopole charge $g$
were found. The expression $f(x)$ for deviation of the solution for
a spheroid from that for a sphere in Eq.~(\ref{eq:16}) was found the
same as for the deviation between spheroidal and spherical
fullerenes in expression (22). So we can expect a similar shift of
the spectra for real elliptically deformed fullerenes when the
magnetic monopole field has to be included inside the surface to
simulate pentagon defects and create fullerenes. Zero energy modes
dictated by the charge $g$ were found for fullerenes contrary to
spherical geometry without the monopole field inside. The shift of
the electronic spectra of spheroidal, in contrast to spherical
fullerenes, gives rise to reduction or increase in the conduction
bandwidth depending on the type of elliptical deformation ($a<c$,
$a>c$). Due to this, the crystals made of these deformed fullerene
molecules, when doped should be poorer or better conductors than the
spherical ones. Moreover, the spherical fullerenes as $C_{60}$ are
stable towards fragmentation than the other bigger
fullerenes~\cite{23}, following our analysis, also with small
elliptical deformations and with a shift of the electronic spectra.
The presence of the magnetic monopole field with the charge leads to
a decrease in the electronic spectra and shifts for spherical and
spheroidal molecules, respectively. The decrease is smaller with
increasing value of the modes of the electronic spectra and,
therefore, for low-lying electronic levels the spectra of the
spheroidal fullerenes $C_{70}$ or C$_{240}$ and C$_{540}$ could be
shifted to a lower magnitude. The very big structures like C$_{960}$
and C$_{1500}$ become more deformed, faceted and can no longer form
a free-electron model like the electronic shell~\cite{18}, which was
the assumption for this model. For these structures the deviation
from the sphericity is larger when the pentagon defects are
localized at the opposite poles. In the case when the poles are far
away from each other we obtain the structure of nanotubes, and for
the exact description some new model related to that proposed here
should be used. We hope that the knowledge of the shifts of the
electronic spectra of spheroids could be useful for experimentalists
for choosing the optimal energetic scale for different types of
fullerenes. Finally, we think that the spheroidal geometry approach
could also be related to other physical problems with slightly
deformed spherical structures that are common in the nature.

\vskip 0.2cm \vskip 0.2cm The author thanks V.A. Osipov for helpful
discussions and advice.
%\vskip 0.4cm \acknowledgments
\vskip 0.2cm \vskip 0.2cm This work was supported by the Slovak
Scientific Grant Agency, Grant No. 3197, the Science and Technology
Assistance Agency under contract  No. APVT-51-027904, and Grant of
Plenipotentiary of Slovakia at JINR.

\end{document}